\documentclass[aps,prl,twocolumn,english,superscriptaddress,longbibliography,nobibnotes]{revtex4-2}
\usepackage{graphicx}
\usepackage[usenames,dvipsnames]{color}
\usepackage{bbm}
\usepackage{bm}
\usepackage{amsmath}
\usepackage{amssymb}
\usepackage{mathptmx}
\usepackage[colorlinks=true,linkcolor=Blue,citecolor=Blue,urlcolor=Blue]{hyperref}
\usepackage[version=3]{mhchem}

\graphicspath{{figures/compressed/}}

\makeatletter
\def\maketitle{
\@author@finish
\title@column\titleblock@produce
\suppressfloats[t]}
\makeatother

\begin{document}

\title{Simulating time evolution on distributed quantum computers}

\author{Finn Lasse Buessen}
\affiliation{Entangled Networks Ltd., Toronto, Ontario M4R 2E4, Canada}
\affiliation{Department of Physics, University of Toronto, Toronto, Ontario M5S 1A7, Canada}
\author{Dvira Segal}
\affiliation{Department of Physics, University of Toronto, Toronto, Ontario M5S 1A7, Canada}
\affiliation{Department of Chemistry and Centre for Quantum Information and Quantum Control, University of Toronto, Toronto, Ontario M5S 3H6, Canada}
\author{Ilia Khait}
\affiliation{Entangled Networks Ltd., Toronto, Ontario M4R 2E4, Canada}

\begin{abstract}
We study a variation of the Trotter-Suzuki decomposition, in which a Hamiltonian exponential is approximated by an ordered product of two-qubit operator exponentials such that the Trotter step size is enhanced for a small number of terms. 
Such decomposition directly reflects hardware constraints of distributed quantum computers, where operations on monolithic quantum devices are fast compared to entanglement distribution across separate nodes using interconnects.
We simulate non-equilibrium dynamics of transverse-field Ising and XY spin chain models and investigate the impact of locally increased Trotter step sizes that are associated with an increasingly sparse use of the quantum interconnect.
We find that the overall quality of the approximation depends smoothly on the local sparsity and that the proliferation of local errors is slow. 
As a consequence, we show that fast local operations on monolithic devices can be leveraged to obtain an overall improved result fidelity even on distributed quantum computers where the use of interconnects is costly.
\end{abstract}

\date{\today}
\maketitle


\begin{figure}[t]
	\includegraphics[width=\linewidth]{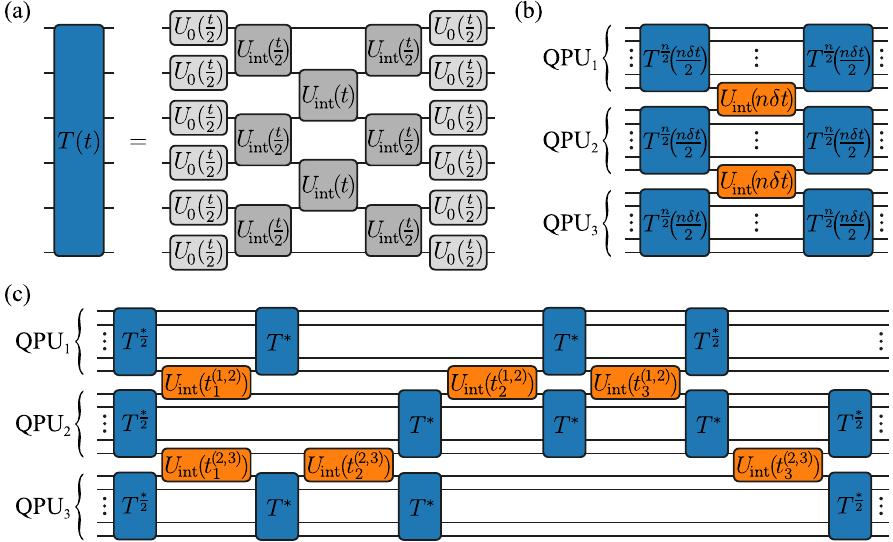}
	\caption{Circuit models for uniform (conventional) and sparse Trotterization schemes on a distributed quantum computer. (a)~Decomposition of a uniform Trotter step $T$ into elementary operations $U_0$ and $U_\mathrm{int}$. (b)~A single Trotter step on a distributed QPU with $k=3$ compute nodes. The sparsity parameter $n$ denotes the number of steps in the node-local Trotter operator (blue) that are performed before a single inter-core operation (orange). (c)~Stochastic sparse Trotterization on $k=3$ distributed compute nodes with step sizes drawn from a random distribution (see text for details).}
	\label{fig:circuits}
\end{figure}

{\it Introduction --}
The availability and rapid evolution of general purpose quantum computing hardware has lead to repeated claims of quantum supremacy~\cite{Arute2019,Zhong2020,Wu2021,Madsen2022}. 
It gives new impetus to Feynman's idea of using quantum computers as platforms to efficiently simulate the dynamics of quantum matter~\cite{Feynman1982}. 
With quantum computers capable enough, the list of potential applications in physics and beyond is long; ranging from condensed matter physics~\cite{Raeisi2012,Macridin2018,Smith2019,Lin2021} to the simulation of general quantum field theories~\cite{Jordan2012,Jordan2014} and from nuclear physics~\cite{Dumitrescu2018} to quantum chemistry~\cite{Wecker2014,Babbush2015,Poulicn2015,Cao2019} and drug discovery~\cite{Cao2018}. 

While quantum supremacy has indeed been claimed for select synthetic problems, current noisy intermediate-scale quantum (NISQ) devices do not yet exceed the performance of classical computers for purposeful algorithms like quantum simulation in the spirit of Feynman~\cite{Preskill2018,Daley2022}. 
Applications on NISQ devices~\cite{Bharti2022} are typically hindered by finite coherence times and insufficient gate precision, as well as by the overall small number of available qubits. 
A crucial step in surpassing the size limitation of current-generation quantum computers will be the transition to distributed quantum computers~\cite{Cuomo2020,VanMeter2016}.
Similar to the limitations of monolithic NISQ devices, however, near-term interconnect hardware that is required for the facilitation of quantum gates across distributed quantum processing units (QPUs, c.f. Fig.~\ref{fig:circuits}) is also facing challenges. 
Most notably, current interconnect hardware generates entanglement between remote qubits at a rate of approximately $182$~Hz, which is over an order of magnitude slower than single-QPU gate operations~\cite{Stephenson2020}. 
It is therefore of great importance in quantum algorithm design to limit the number of required interconnect uses. 

In this work, we study the Trotter-Suzuki decomposition~\cite{Suzuki1991,Trotter1959,Suzuki1976} from the perspective of potential implementations on distributed quantum hardware. 
The Trotter-Suzuki decomposition is a common technique for approximating many-body time evolution operators by a product of two-qubit operators that can be implemented on NISQ devices. 
We propose a variation of the decomposition, in which the resulting two-qubit operators are no longer treated on equal footing: 
To limit the number of inter-core operations, two-qubit gates that are facilitated by a quantum interconnect are treated at a coarser level of approximation, while a more accurate approximation, i.e., a shorter Trotter step size, is maintained for operations that are local within a single compute node. 
Our variation of the decomposition, which is illustrated in Fig.~\ref{fig:circuits}, allows us to systematically control the sparsity of interconnect usage. 

For select models of non-equilibrium quantum magnetism, we utilize our approach to reduce the number of interconnect uses. 
We demonstrate that a significant enhancement of accuracy is achieved over the traditional Trotter-Suzuki decomposition when the Trotter step size is limited by the interconnect rate. 
Finally, we also benchmark a scenario in which the sparsity of interconnect usage is randomized.
This emulates the operation of an interconnect at its latency limit, when due to the non-deterministic nature of its dead-time after each use it cannot be guaranteed that the link is immediately available~\cite{Olmschenk2009}.


{\it Models --}
We consider two distinct models of quantum magnetism. 
Each model is based on a one-dimensional finite chain of $L$ spin-1/2 operators that are represented by the three Pauli matrices $(\sigma^x_i, \sigma^y_i, \sigma^z_i)$, where $i$ denotes the position on the chain.
The first model is the XY model, governed by the Hamiltonian $H_\mathrm{XY} = -J \sum_{\langle i,j \rangle} ( \sigma_i^x\sigma_j^x + \sigma_i^y\sigma_j^y )$, where $\langle ~,~ \rangle$ denotes nearest neighbor pairs on the chain and we fix the interaction energy scale $J=1$.
For the time evolution under this model, we assume that the system is initially prepared in a domain wall state $| \uparrow\dots\uparrow\downarrow\dots\downarrow \rangle$ in the eigenbasis of $\sigma^z$. 
The second model is the transverse-field Ising (TFI) model with Hamiltonian $H_\mathrm{TFI} = -J \sum_{\langle i,j \rangle} \sigma_i^z\sigma_j^z + h \sum_i \sigma^x_i$. 
We consider the case $h=0.5$, which we refer to as a \emph{slow} quench, and $h=2.0$, which we call a \emph{fast} quench~\cite{Pfeuty1970,Calabrese2012}. 
In both cases for the TFI model, we assume the spin chain to initially be uniformly ordered with all spins in the $|\downarrow\rangle$ state. 

In the following, we assume that spins are represented by qubits, and we refer to them synonymously. 
For quantum simulation, the spin chain is divided into $k$ sections of equal size and mapped onto $k$ quantum compute nodes that are interconnected linearly, cf. Fig.~\ref{fig:circuits}.


{\it Sparse Trotterization --}
The model Hamiltonians outlined in the previous section can be summarized in the generalized notation $H=\sum_i H_0^{(i)} + \sum_{\langle i,j \rangle} H_\mathrm{int}^{(i,j)}$, where $H_0^{(i)}$ denotes a local term on site $i$ and $H_\mathrm{int}^{(i,j)}$ denotes an interaction term between neighboring sites $i$ and $j$; the different terms generally do not commute. 
The time evolution operator $U(t)=e^{-iHt}$ can be approximated by a sequence $T^N(t)$ of one- and two-qubit operations $U_0^{(i)}(t) = e^{-i H_0^{(i)} t}$ and $U_\mathrm{int}^{(i,j)}(t) = e^{-i H_\mathrm{int}^{(i,j)} t}$ as
\begin{equation}
\label{eq:trotterization:uniform}
T^N(t) = \left( T_0(\tfrac{\delta t}{2}) T_\mathrm{even}(\tfrac{\delta t}{2}) T_\mathrm{odd}(\delta t) T_\mathrm{even}(\tfrac{\delta t}{2}) T_0(\tfrac{\delta t}{2}) \right)^N \,,
\end{equation}
where $\delta t = \frac{t}{N}$, $T_0(t) = \prod_i U_0^{(i)}(t)$, $T_\mathrm{even}(t) = \prod_{\langle i,j \rangle_\mathrm{even}} U_\mathrm{int}^{(i,j)}(t)$ with the product running over all even pairs of nearest neighbor sites (nearest neighbor pairs on the spin chain are alternatingly labeled as even and odd) and $T_\mathrm{odd}(t)$ defined analogously. 
This approximation is known as the ($N$-step) second-order Trotter-Suzuki decomposition, depicted in Fig.~\ref{fig:circuits}a for a single step ($N=1$)~\cite{Suzuki1991, Trotter1959}.
The approximation error scales $\sim N \delta t^3$~\footnote{In some cases, tighter bounds can be formulated, depending on the structure of the underlying Hamiltonian~\cite{Layden2022,Childs2021,Heyl2019,Sieberer2019}.}. 
For the remainder of the manuscript, we shall also refer to this approximation as \emph{uniform} Trotterization. 

Motivated by the expected constraints for near-term distributed quantum computing architectures -- most importantly, the slow rate of entanglement generation in quantum interconnects -- we define \emph{sparse} Trotterization as follows. 
We assume a distributed quantum computer to consist of $k>1$ compute nodes that are interconnected linearly, cf.~Fig.~\ref{fig:circuits}.
Interconnects may only be used at a fraction $1/n$ of the speed at which each individual compute node operates; we refer to $n$ as the \emph{sparsity}. 
The sparse Trotterization of the time evolution operator $U(t)$ is defined as
\begin{equation}
\label{eq:trotterization:sparse}
T_\mathrm{sparse}^{N,n}(t) \!=\!\! \Bigg( \!\prod\limits_{\kappa=1}^k \!T^{\frac{n}{2}}\!\Big|_\kappa\!\!(\!\tfrac{n\delta t}{2}\!) \!\!\!\prod\limits_{\langle i,j \rangle_{\kappa\neq\kappa'}} \!\!\!\!\! U_\mathrm{int}^{(i,j)}\!(n \delta t) \prod\limits_{\kappa=1}^k \!T^{\frac{n}{2}}\!\Big|_\kappa\!\!(\!\tfrac{n\delta t}{2}\!) \!\!\! \Bigg)^{\!\!\!\!\!\frac{N}{n}} \!\!\!\!,
\end{equation}
where $T|_\kappa(t)$ denotes the usual (uniformly Trotterized) time evolution within a single compute node $\kappa$ and $\langle i,j \rangle_{\kappa\neq\kappa'}$ denotes nearest neighbor pairs of qubits $i$ and $j$ on separate compute nodes $\kappa$ and $\kappa'$. 
By virtue of this definition, the step size within each compute node remains $\delta t = \frac{t}{N}$, yet the interaction between qubits on different compute nodes is computed with larger step size $n \delta t$. 
The corresponding circuit model is depicted in Fig.~\ref{fig:circuits}b for one sparse Trotter step with a single interconnect use. 

We further define a \emph{stochastic} sparse Trotterization, in which the time steps for remote operations are randomized. 
The definition is as follows: 
For each interconnect between sites $i$ and $j$ on compute nodes $\kappa$ and $\kappa'$, randomly choose time intervals $t^{(\kappa,\kappa')}_1,t^{(\kappa,\kappa')}_2,\dots$ such that $t^{(\kappa,\kappa')}_1 + t^{(\kappa,\kappa')}_2 + \dots = t$. 
Next, apply remote operations $U_\mathrm{int}^{(i,j)}(t^{(\kappa,\kappa')}_1),U_\mathrm{int}^{(i,j)}(t^{(\kappa,\kappa')}_2),\dots$ on every interconnect. 
Before every such remote operation, insert local time evolutions $T^*|_\kappa$ and $T^*|_{\kappa'}$ with $T^*\equiv T^{N^*}(t^*)$ and parameters $N^*$ and $t^*$ such that the total time evolved locally with the usual step size $\delta t$ matches the total time evolved on the interconnect with coarser step size $t^{(\kappa,\kappa')}_1,t^{(\kappa,\kappa')}_2,\dots$~\cite{SM}. 
An example of the resulting circuit model is displayed in Fig.~\ref{fig:circuits}c. 
This variation of the Trotter-Suzuki decomposition is intended to reflect the limitations of interconnects, which generate entanglement non-deterministically. 
It allows to stretch the duration of the node-local time evolution -- and thus the number of gates applied and the absolute computing time -- until the required interconnect becomes available. 
We note that inhomogeneous variations of the Trotter-Suzuki decomposition at various orders~\cite{Papageorgiou2012,Berry2007} have been studied previously in the context of quantum chemistry, where different terms in the electronic Hamiltonian are separated by their energy scale to allow for a reduction of the Trotter step size within controlled error bounds~\cite{Hadfield2018,Childs2019,Campbell2019,Ouyang2020}. 
Here, our motivation to consider inhomogeneous step sizes is rooted in hardware constraints of a distributed quantum computing platform and separation occurs according to qubit connectivity.


\begin{figure*}
	\includegraphics[width=\linewidth]{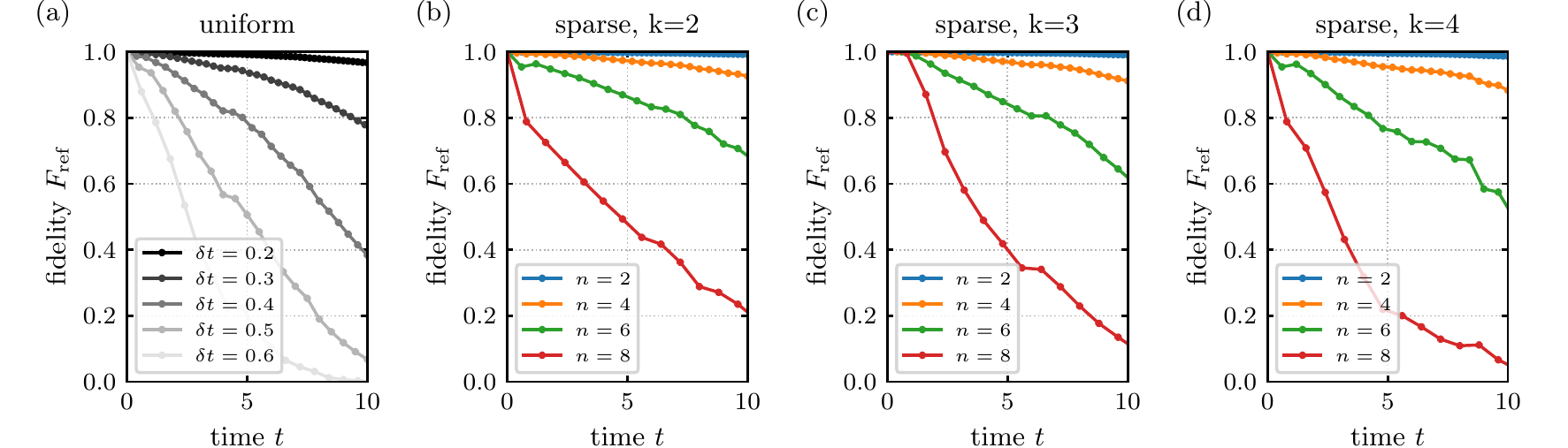}
	\caption{Sparse Trotterization for the XY model with an initial domain wall configuration at the chain center. (a)~Fidelity $F_\mathrm{ref}$ obtained for the time evolution under uniform Trotterization with various step sizes $\delta t$. The fidelity is computed with respect to the reference state obtained from step size $\delta t_\mathrm{ref}=0.1$. (b)--(d)~Fidelity of states obtained from sparse Trotterization across $k=2,3$, and $4$ distributed quantum compute nodes at different levels of sparsity $n$. The node-local step size is $\delta t_\mathrm{ref}=0.1$. Data is obtained for $L=24$ spins.}
	\label{fig:xy_domainwall_fidelity}
\end{figure*}
\begin{figure}
	\includegraphics[width=\linewidth]{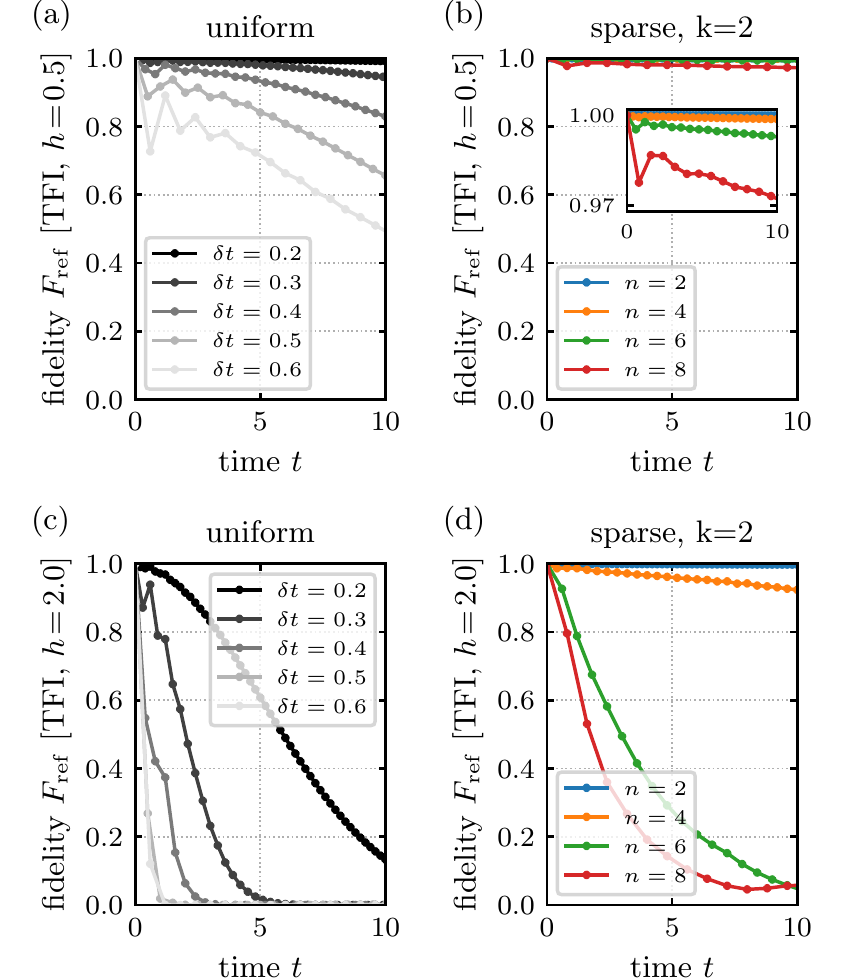}
	\caption{Sparse Trotterization for TFI quench models. (a)~Fidelity $F_\mathrm{ref}$ for uniform Trotterization with various step sizes $\delta t$, obtained for the slow TFI quench model. The fidelity is computed with respect to the reference state obtained from step size $\delta t_\mathrm{ref}=0.1$. (b)~Fidelity obtained from sparse Trotterization at varying sparsity $n$ on $k=2$ distributed quantum compute nodes. The node-local step size is $\delta t_\mathrm{ref}=0.1$. (c)--(d)~Same as panels (a)--(b) but for the fast TFI quench model. Data is obtained for $L=24$ spins.}
	\label{fig:tfi_uniform_fidelity}
\end{figure}

{\it Results --}
We begin our analysis by investigating the role of sparsity for the example of the XY model and compare its performance to the conventional uniform Trotterization. 
In the following, we shall assume a reference step size of $\delta t_\mathrm{ref} =0.1$ for node-local operations. 
Note that for sparse Trotterization, a sparsity of $n$ would then entail that remote operations are performed with a step size of $n\delta t_\mathrm{ref}$. 
We evaluate all approximations by computing the associated wave functions and benchmarking against a reference state $|\psi_\mathrm{ref}(t)\rangle$ that is obtained from uniform Trotterization with time step $\delta t_\mathrm{ref}$~\cite{SM}. 
The quality of any state $|\psi(t)\rangle$ can then be quantified by the reference fidelity $F_\mathrm{ref}=\left| \langle \psi_\mathrm{ref}(t) | \psi(t) \rangle \right|^2$. 
We find that for increased step size $\delta t=0.2$ in a uniform Trotterization the fidelity remains acceptable with $F_\mathrm{ref}=0.97$ after evolution to $t=10$ and diminishes quickly for larger step sizes, see Fig.~\ref{fig:xy_domainwall_fidelity}a. 
In contrast, on a distributed architecture with $k=2$ compute nodes, introducing a sparsity of $n=2$ still yields a fidelity of $F_\mathrm{ref}=0.99$ after evolution to $t=10$ and the decay in result quality for increasing $n$ is reduced significantly (Fig.~\ref{fig:xy_domainwall_fidelity}b):
Sparse Trotterization with $n=4$ ($F_\mathrm{ref}=0.93$) still performs better than uniform Trotterization with $\delta t=0.3$ ($F_\mathrm{ref}=0.77$), despite the larger step size for the interconnect-mediated interaction.
The trend not only holds for larger $n$, but also as the number of compute nodes $k$ is increased moderately, see Figs.~\ref{fig:xy_domainwall_fidelity}c and~\ref{fig:xy_domainwall_fidelity}d. 

We make similar observations for the slow TFI quench model, where the fidelity is significantly more robust against local sparsity $n$ than against a globally increased step size (Figs.~\ref{fig:tfi_uniform_fidelity}a, \ref{fig:tfi_uniform_fidelity}b).
The robustness may be related to the finite magnetization in the initial state, which persists well beyond $t=10$ and can act self-stabilizing against local perturbations~\cite{SM}. 
In the fast TFI quench model, the fidelity decreases rapidly for $n\geq 6$.
Yet, this still marks a substantial improvement over a global increase of the step size $\delta t$, especially for small $n=2$ and $n=4$ (Figs.~\ref{fig:tfi_uniform_fidelity}c, \ref{fig:tfi_uniform_fidelity}d). 

For practical applications, usually the goal is to accurately predict physical observables like the time-dependent magnetization $m_i^z(t)=\langle \psi(t) | \sigma_i^z | \psi(t)\rangle$ or the magnetic correlation function $\chi_{i,j}^{zz}(t)=\langle \psi(t)|\sigma_i^z\sigma_j^z |\psi(t)\rangle$. 
For the XY model on $k=2$ compute nodes, the magnetization and magnetic correlations obtained at sparsity $n=4$ are depicted in Figs.~\ref{fig:xy_domainwall_mag_corr}a and \ref{fig:xy_domainwall_mag_corr}c, witnessing oscillatory behavior during the decay of the domain wall. 
Deviations from the magnetization and correlations of the reference state $|\psi_\mathrm{ref}(t)\rangle$ are small, especially when compared to the error that accumulates when the step size is uniformly increased in a uniform Trotterization. 
The deviation for the magnetization at the chain boundary, $m_0^z(t)$, is illustrated in Fig.~\ref{fig:xy_domainwall_mag_corr}b and for the correlation between the chain boundary and the bulk, $\chi_{0,5}^{zz}(t)$, in Fig.~\ref{fig:xy_domainwall_mag_corr}d. 
More generally, we compute deviations from the reference magnetization as $\Delta m_i^z(t)=m_i^z(t) - \langle \psi_\mathrm{ref}(t) | \sigma_i^z | \psi_\mathrm{ref}(t)\rangle$, see the Supplemental Material (SM) for the full space- and time-resolved data~\cite{SM}. 
The maximum deviation $\max_{i,t} (|\Delta m_i^z(t)|)$ obtained with sparsity $n=(2,4,6)$ is $(0.03, 0.10, 0.22)$.
In contrast, the maximum deviation for states obtained from uniform Trotterization with analogous uniform step sizes $\delta t=(0.2,0.4,0.6)$ is significantly larger, yielding $(0.15, 0.58, 0.94)$. 
Differences of similar magnitude are also observed for the deviation of the magnetic correlations~\cite{SM}.

\begin{figure}
	\includegraphics[width=\linewidth]{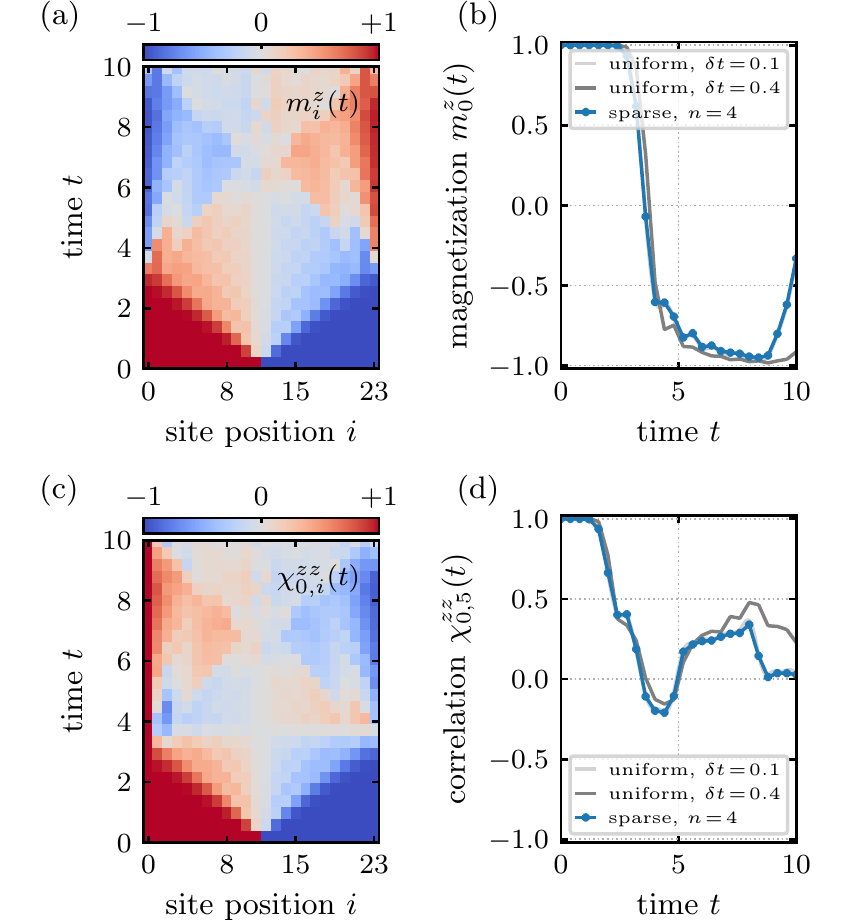}
	\caption{Local magnetization and magnetic correlations in the XY domain wall model. (a)~Time-dependent magnetization obtained with sparsity $n=4$. (b)~Local magnetization at the boundary of the chain, $m_0^z(t)$, obtained from sparse Trotterization with $n=4$ and uniform Trotterization with $\delta t=0.1, 0.4$. The curve for sparse Trotterization coincides with the curve for uniform Trotterization at $\delta t=0.1$. (c)--(d)~Same as panels (a)--(b) but for the magnetic correlation function. Data is computed for $L=24$ spins across $k=2$ compute nodes and a node-local step size $\delta t_\mathrm{ref}=0.1$.}
	\label{fig:xy_domainwall_mag_corr}
\end{figure}

\begin{figure}[b]
	\includegraphics[width=\linewidth]{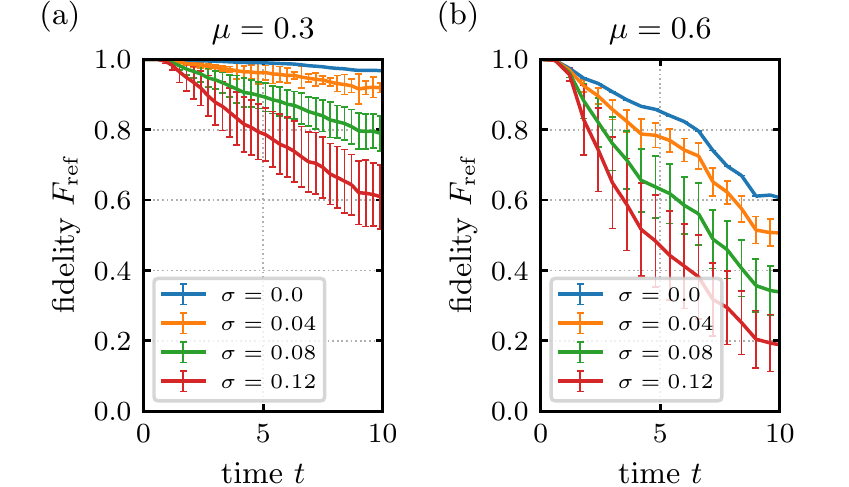}
	\caption{Stochastic sparse Trotterization of the time evolution for the XY model at $k=3$. The fidelity is shown for different levels of standard deviation~$\sigma$. Average sparsity is set to (a)~$\mu=0.3$ and (b)~$\mu=0.6$, respectively. Data is obtained for $L=18$ spins and node-local step size $\delta t_\mathrm{ref}=0.1$.}
	\label{fig:xy_domainwall_randomized}
\end{figure}

The above results indicate that the quality of the sparse Trotterization, for the Hamiltonians with two-spin interactions considered in this work, smoothly depends on the sparsity parameter $n$ and on the number of sparse qubit pairs $k-1$. 
Increasing the Trotter step size only between a small number of $k-1$ qubit pairs does not immediately lead to a proliferation of the error to levels that are associated with a uniform increase of $\delta t$ between all qubit pairs. 
Therefore, the results of a conventional Trotter-Suzuki decomposition with a given step size $\delta t_\mathrm{uniform}$ can be matched or improved by an inhomogeneous decomposition for which the step size is more fine-grained between most of the qubit pairs ($\delta t < \delta t_\mathrm{uniform}$) but coarser between a small number of qubit pairs ($\delta t > \delta t_\mathrm{uniform}$). 
For the models studied here, a reduction of the number of trotter steps between a small number of qubit pairs by about 50\% seems feasible.

Finally, we explore the effect of randomness in the stochastic sparse Trotterization. 
For this purpose, the time steps on the sparse qubit pairs are drawn from a normal distribution with mean $\mu$ and standard deviation $\sigma$. 
We then calculate $F_\mathrm{ref}$ and average it over 1000 instances of randomized configurations for each parameter set.
In practical applications, stochastic averaging occurs naturally when measurement results are sampled statistically and interconnect uses on distributed quantum computing systems are inherently randomized.
We find that despite relatively large variations across the different randomized configurations, the mean value for the fidelity remains smooth and systematically depends on $\mu$ and $\sigma$, see Fig.~\ref{fig:xy_domainwall_randomized} for data on the XY model on $k=3$ distributed compute nodes.
Data for the TFI quench models is shown in the SM~\cite{SM}. 
Our data also indicates that randomness $\sigma$ can have a more adverse effect than a systematic increase of the mean step size $\mu$. 
For example, $F_\mathrm{ref}$ at $(\mu,\sigma)=(0.3,0.12)$ is comparable to the fidelity achieved for much larger mean step size in the absence of randomness, $(\mu,\sigma)=(0.6, 0.0)$, see Fig.~\ref{fig:xy_domainwall_randomized}. 
Note that in the former case a step size of less than $0.6$ would effectively occur with probability greater than 99\% and a fidelity enhancement would therefore naively be expected. 
We speculate that the randomness leads to a reduction in the cancellation of Trotterization error terms that has been observed to be relevant for practical models in condensed matter physics~\cite{Layden2022,Sieberer2019,Heyl2019}.


{\it Conclusion --}
We have demonstrated that issues in the implementation of quantum simulations on distributed quantum computers that arise from slow interconnect hardware can be mitigated by modifying the Trotter-Suzuki decomposition to allow for a non-uniform variation of the Trotter step size. 
For the XY model and the TFI quench models studied in this letter, a coarsening of the step size between a small number of qubit pairs could be compensated by a refinement of the step size across the remaining qubit pairs. 
Notably, the sparse Trotterization could be applied successfully despite the fact that the underlying models only have a single principal energy scale and therefore ruling out Trotter constructions that rely on scale separation~\cite{Hadfield2018,Childs2019,Campbell2019,Ouyang2020}.  
This observation has important consequences for the implementation of quantum simulations on distributed quantum computers. 
In current hardware, individual compute nodes of a distributed quantum computer operate at significantly faster gate speed compared to the interconnect between different compute nodes. 
Future hardware generations are expected to yield higher fidelities and increased interconnect speed, ultimately allowing for a greater number of Trotter steps to be executed. 
Yet, gate operations that are facilitated by an interconnect are expected to remain slower than operations within a single node. 
To remedy the speed deficiency, instead of using Trotterization with a uniform time step that is bounded by the interconnect speed, results of similar or better fidelity can be obtained by maintaining fine-grained time stepping within each compute node and using coarser time steps for remote operations. 
For the examples considered here, we find that a reduction in the number of interconnect uses by as much as 50\% can be viable. 

Further, we explored the possibility of exploiting the non-deterministic dead time after every interconnect use to perform additional node-local Trotter steps until the interconnect becomes available. 
Our data suggests that the randomized execution of additional Trotter steps quickly degrades the result quality. 
Unless the overall reduction in total compute time can compensate for the randomness-induced fidelity loss, it remains more beneficial to delay the execution of additional Trotter steps. 
Further calculations with hardware specific error models are required to find the optimal tradeoff. 

In this work, we focused on one-dimensional models. 
To explore more general applications in the future, it would be interesting to benchmark the performance of sparse Trotterization for quantum spin models with next-nearest neighbor interactions or beyond, as well as for models in higher dimensions. 
In such generalizations, increased inter-node communication is expected and additional optimization of the interconnect usage may be necessary~\cite{Tham2022, MultiQopt2022}.
Further practical applications could also include the sparse Trotterization of imaginary time evolution~\cite{Motta2020} or the integration with variational algorithms~\cite{Lin2021} on distributed quantum computers.


\begin{acknowledgments}
This work was supported by Mitacs through the Mitacs Elevate program.
DS acknowledges the NSERC discovery grant and the Canada Research Chair Program. 
\end{acknowledgments}


\bibliography{trotterization}


\clearpage
\renewcommand{\thefigure}{S\arabic{figure}}
\renewcommand{\theHfigure}{S\arabic{figure}}
\setcounter{figure}{0}

\title{Supplementary information: Simulating time evolution on distributed quantum computers}
\maketitle
\onecolumngrid

\section{Stochastic sparse Trotterization}
In this section, we provide additional information on the definition of the stochastic sparse Trotterization. 
In the following, we shall assume that random numbers are drawn from a normal distribution with mean value~$\mu$ and standard deviation~$\sigma$. 
We further assume the first time step $t_1$ to be deterministic, since any quantum interconnect can be readily initialized before the start of a calculation. 
In our notation, the time evolution begins at $t=0$ and ends at $t_\mathrm{end}$. 
The uniform Trotterization within each compute node is performed with a reference step size $\delta t_\mathrm{ref}$. 

\begin{figure}
	\includegraphics[width=\linewidth]{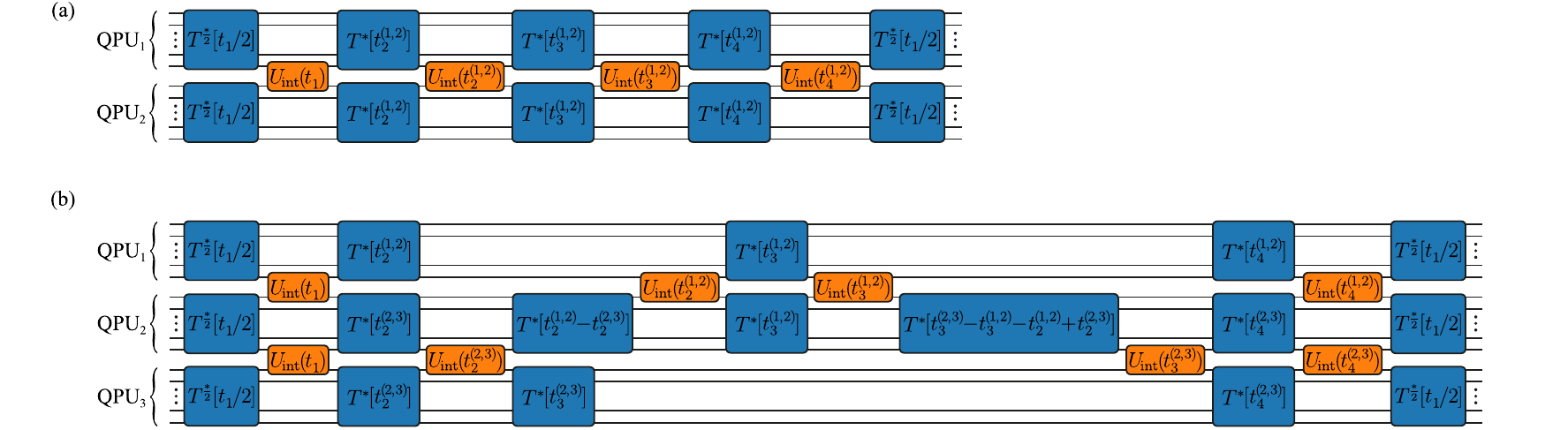}
	\caption{Instances of stochastic sparse Trotterization in circuit notation for (a)~$k=2$ compute nodes and (b)~$k=3$ compute nodes. Blue gates denote uniform Trotterization within a single compute node. Orange gates denote time evolution that operates on qubits across two distinct compute nodes, see text for details. Note that before application of the gate $U_\mathrm{int}(t_4^{(1,2)})$ in panel~(b), the total time evolved on QPU$_1$ is $t_1/2+t_1^{(1,2)}+t_2^{(1,2)}+t_3^{(1,2)}+t_4^{(1,2)}$ and the total time evolved on QPU$_2$ is $t_1/2+t_1^{(2,3)}+t_2^{(2,3)}+t_3^{(2,3)}+t_4^{(2,3)}$; since both are equal to $t_\mathrm{end}-t_1/2$, no additional Trotterization operation needs to be inserted.}
	\label{fig:circuits_extended}
\end{figure}

For $k=2$ interconnected compute nodes, the definition of the stochastic sparse Trotterization is straight-forward: 
Start by drawing sufficiently many random numbers $t_2^{(1,2)},t_3^{(1,2)},\dots$ such that $t_1 + t_2^{(1,2)} + t_3^{(1,2)} + \dots \geq t_\mathrm{end}$. 
In drawing the random numbers, we shall assume a minimum value of $\delta t_\mathrm{ref}$; any random number smaller than $\delta t_\mathrm{ref}$ shall be replaced by $\delta t_\mathrm{ref}$. 
Furthermore, the last random number shall be replaced by a value such that the equality $t_1 + t_2^{(1,2)} + t_3^{(1,2)} + \dots = t_\mathrm{end}$ holds exactly.
To construct the circuit representation of the stochastic sparse Trotterization, begin by inserting an operation $T^{\frac{*}{2}}[t_1/2]$ on every compute node. 
The operation is defined as a uniform Trotterization for a total time of $t_1/2$ with an appropriate number of Trotter steps of reference step size $\delta t_\mathrm{ref}$. 
If $t_1/2$ is not evenly divisible by $\delta t_\mathrm{ref}$, we shall perform $\lfloor t_1/(2\delta t_\mathrm{ref}) \rfloor$ Trotter steps of reference step size $\delta t_\mathrm{ref}$ and one step of step size $t_1/2 - \lfloor t_1/(2\delta t_\mathrm{ref}) \rfloor \delta t_\mathrm{ref}$, where $\lfloor . \rfloor$ denotes rounding to the next lower integer. 
We then append the operation $U_\mathrm{int}(t_1)$ on the sparsified bond, i.e., the bond that connects qubits across the two compute nodes.
A second operation $T^{\frac{*}{2}}[t_1/2]$ is appended to the end of the circuit on every compute node. 
This construction of symmetrizing the first Trotter step into two half-steps ensures that in the absence of any randomness the circuit reduces to the definition of the sparse Trotterization. 
In the bulk part of the circuit, i.e., after the operation $U_\mathrm{int}(t_1)$, we now insert all remaining operations. 
This is done by successively inserting, for $i=2,3,\dots$, node-local uniform Trotterizations $T^*[t_i^{(1,2)}]$ on every node followed by an operation $U_\mathrm{int}(t_i^{(1,2)})$ on the sparse bond. 
In analogy to the first Trotter step, $T^*[t_i^{(1,2)}]$ denotes the insertion of an appropriate number of Trotter steps with reference step size $\delta t_\mathrm{ref}$. 
An example circuit for random time steps $t_1, t_2^{(1,2)}, t_3^{(1,2)}, t_4^{(1,2)}$ is displayed in Fig.~\ref{fig:circuits_extended}a.

For $k=3$ interconnected compute nodes, we extend the definition as follows. 
For the sparse bond between compute nodes~1 and~2, generate the sequence of time steps $t_2^{(1,2)}, t_3^{(1,2)}, \dots$ as previously defined. 
Similarly, generate a sequence of time arguments $t_2^{(2,3)}, t_3^{(2,3)}, \dots$ for Trotter steps between compute nodes~2 and~3. 
The first and last circuit layers with node-local Trotterization $T^{\frac{*}{2}}[t_1/2]$ and time evolution $U_\mathrm{int}(t_1)$ on the sparsified bonds is performed in analogy to the definition for $k=2$. 
In the bulk part of the circuit we proceed as follows. 
All time steps $t_i^{(m,n)}$ are ordered according to a key $t_1/2+\sum_{j=2}^i t_j^{(m,n)}$ in ascending order, i.e., they are ordered by the total time evolved on the respective interconnect up to the time step $t_i^{(m,n)}$. 
We then iteratively take the first value $t_i^{(m,n)}$ off the ordered list and apply node-local uniform Trotterizations to the compute nodes $m$ and $n$ such that the total time evolved on those nodes matches $t_1/2+\sum_{j=2}^i t_j^{(m,n)}$, followed by the time evolution $U_\mathrm{int}(t_i^{(m,n)})$ on the sparsified bond between compute nodes $m$ and $n$. 
An example of the resulting circuit for $k=3$ is shown in Fig.~\ref{fig:circuits_extended}b. 
The generalization to $k\geq4$ is defined analogously. 

\begin{figure}
	\includegraphics[width=0.5\linewidth]{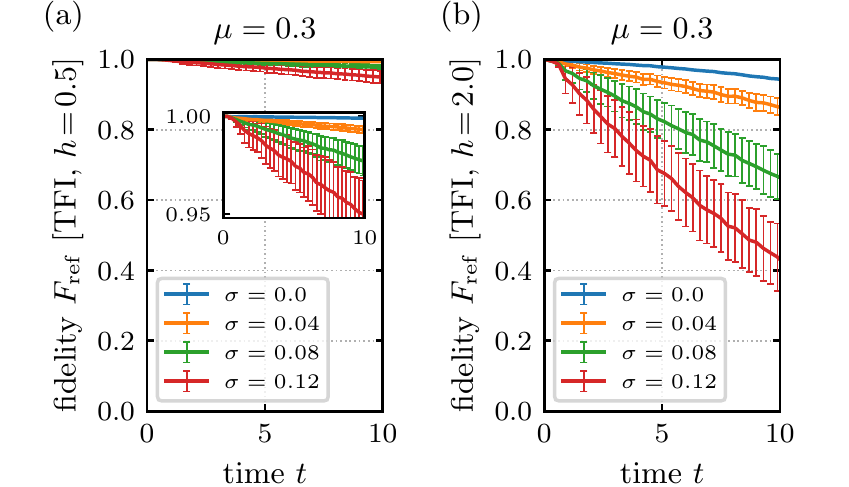}
	\caption{Stochastic sparse Trotterization of the time evolution for (a)~the slow TFI quench model and (b)~the fast TFI quench model at $k=3$. The fidelity is shown for different levels of standard deviation~$\sigma$ at fixed mean sparsity $\mu=0.3$. Data is for $L=18$ spins and node-local step size $\delta t_\mathrm{ref}=0.1$.}
	\label{fig:tfi_uniform_randomized}
\end{figure}

The reference fidelity $F_\mathrm{ref}$ with respect to the state obtained from uniform Trotterization with reference step size $\delta t_\mathrm{ref}=0.1$ is shown for the TFI quench models on $k=3$ compute nodes in Fig.~\ref{fig:tfi_uniform_randomized}. 
The data has been obtained by averaging over 1000 time step configurations, where the time steps were randomly drawn from a normal distribution with mean $\mu=0.3$ and different values for the standard deviation $\sigma=0.0$, $0.04$, $0.08$, $0.12$. 

\section{Reference states}
\begin{figure}[b]
	\includegraphics[width=\linewidth]{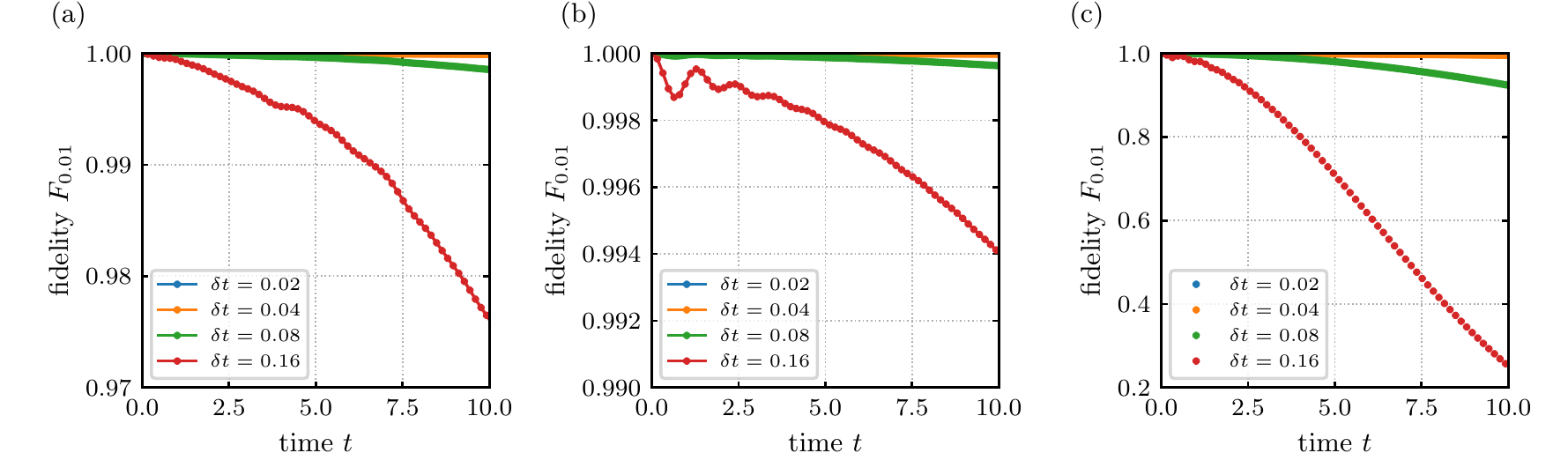}
	\caption{Fidelity $F_{0.01}$ of states obtained from uniform Trotterization with various step sizes $\delta t$, measured with respect to the state obtained from $\delta t_\mathrm{ref}' = 0.01$. Data shown is for (a)~the XY model, (b)~the slow TFI quench model with $h=0.5$, and (c)~the fast TFI quench model with $h=2.0$. The curves for $\delta t=0.02$ and $\delta t=0.04$ are indiscernible in all subpanels. Data is obtained for $L=24$ spins.}
	\label{fig:reference_fidelity}
\end{figure}

In the main manuscript, we measure all fidelities $F_\mathrm{ref}=\left| \langle \psi_\mathrm{ref}(t) | \psi(t) \rangle \right|^2$ with respect to the state $\psi_\mathrm{ref}(t)$ obtained from a uniform Trotterization with step size $\delta t_\mathrm{ref}=0.1$. 
Previous calculations on currently available quantum hardware have demonstrated that around five Trotter steps can be achieved~\cite{Smith2019}. 
We chose the reference step size $\delta t_\mathrm{ref}=0.1$ since it seems achievable on near-term hardware for reaching total time evolution until relevant time scales of around $t\approx 2$, requiring a total of 20 Trotter steps. 

At the same time, the state $\psi_\mathrm{ref}(t)$ is reasonably close to the exact solution. 
To demonstrate the latter, we compute the fidelity $F_{0.01}$ with respect to the state obtained via a much finer time step $\delta t_\mathrm{ref}'=0.01$ for uniform Trotterization with various step sizes $\delta t = (0.02, 0.04, 0.08, 0.16)$. 
For the XY model, the coarsest step size $\delta t = 0.16$ already achieves a fidelity $F_{0.01}>0.97$, see Fig.~\ref{fig:reference_fidelity}a. 
The solution for the slow TFI quench model reaches $F_{0.01}>0.99$ at the same step size (Fig.~\ref{fig:reference_fidelity}b). 
Only the state obtained for the fast TFI quench model has substantially lower fidelity, underscoring the difficulty of accurately simulating the model. 
For $\delta t = 0.16$, the fidelity is reduced to $F_{0.01}=0.24$, but reaching substantially higher values for $\delta t = 0.08$ with $F_{0.01}=0.92$ (Fig.~\ref{fig:reference_fidelity}c).

\section{Magnetization and magnetic correlations}
Here, we show additional data on the accuracy at which physical observables are computed within the sparse Trotterization scheme. 
For this purpose, we compute the time-dependent magnetization $m_i^z(t)=\langle \psi(t) | \sigma_i^z | \psi(t)\rangle$ and magnetic correlation function $\chi_{i,j}^{zz}(t)=\langle \psi(t)|\sigma_i^z\sigma_j^z |\psi(t)\rangle$ for different values of sparsity $n$. 
We then determine the deviation $\Delta m_i^z(t) = m_i^z(t) - \langle \psi_\mathrm{ref}(t) | \sigma_i^z | \psi_\mathrm{ref}(t)\rangle$ from the magnetization of the reference state $| \psi_\mathrm{ref}(t)\rangle$ that is obtained from uniform Trotterization with reference step size $\delta t_\mathrm{ref}=0.1$. 
The deviation of the magnetic correlations, $\Delta \chi_{i,j}^{zz}(t)$, is computed analogously. 

\begin{figure}
	\includegraphics[width=\linewidth]{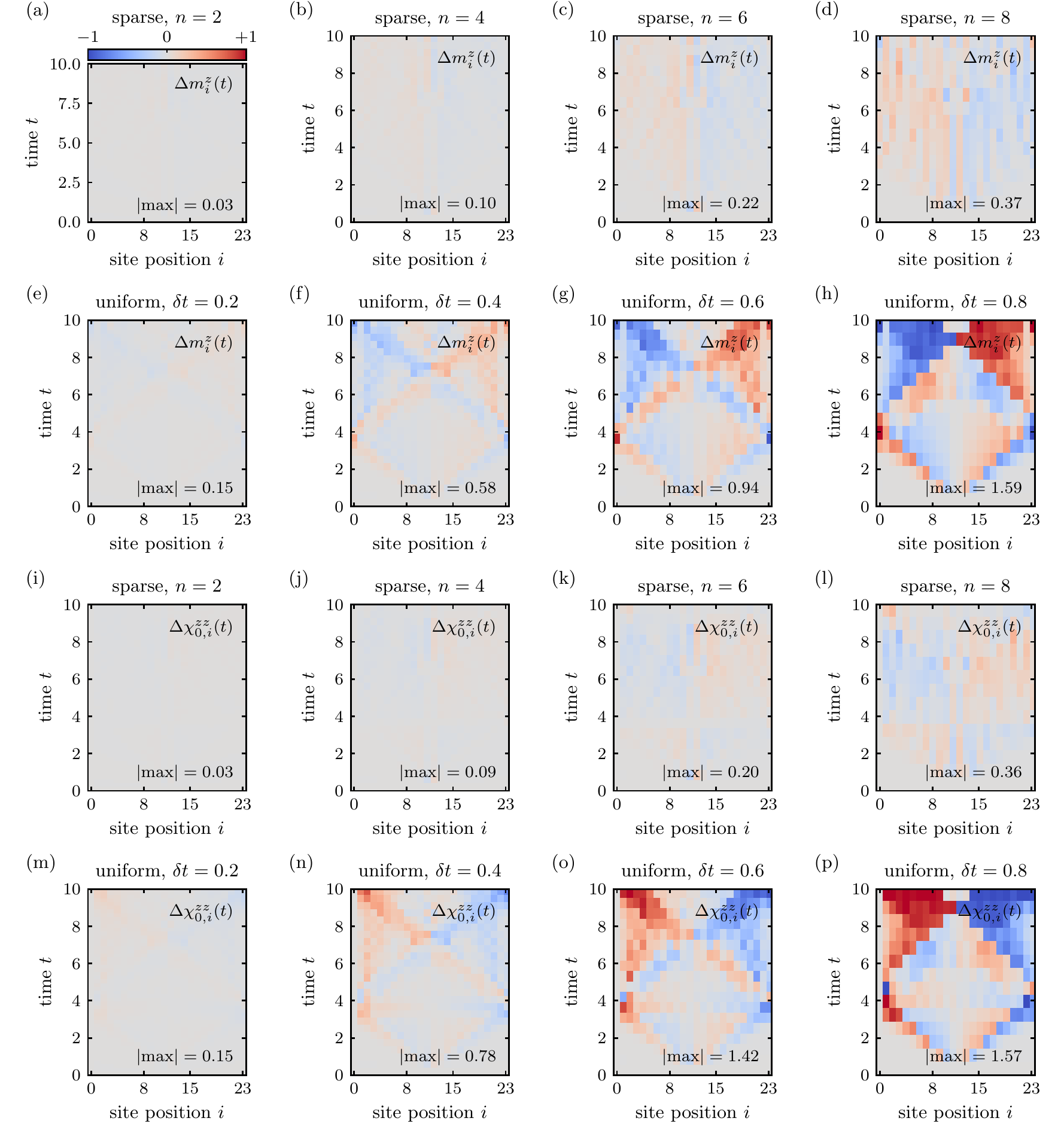}
	\caption{Magnetization and magnetic correlations in the XY model. All data is calculated for two compute nodes, $k=2$. (a)--(d)~Magnetization difference $\Delta m_i^z(t)$ between states obtained from sparse Trotterization with different sparsity $n$ and the reference state $|\psi_\mathrm{ref}(t)\rangle$. The maximum deviation $|\mathrm{max}| = \max_{i,t}(|\Delta m_i^z(t)|)$ across all spins and within the full time range shown is indicated in each panel. (e)--(g)~Magnetization difference between states obtained from uniform Trotterization with different step size $\delta t$ and the reference state. (i)--(l)~Correlation difference $\Delta\chi_{0,i}^{zz}(t)$ between states obtained from sparse Trotterization with different sparsity $n$ and the reference state. The maximum deviation $|\mathrm{max}| = \max_{i,t}(|\Delta\chi_{0,i}^{zz}(t)|)$ is indicated in each panel. (m)--(p)~Correlation difference between states obtained from uniform Trotterization with various step sizes $\delta t$ and the reference state. The color bar shown in panel~(a) applies to all panels. Data is obtained for $L=24$ spins.}
	\label{fig:xy_domainwall_mag_corr_deviation}
\end{figure}
The magnetization deviation for the XY model, obtained for $k=2$ compute nodes, is shown in Figs.~\ref{fig:xy_domainwall_mag_corr_deviation}a--\ref{fig:xy_domainwall_mag_corr_deviation}d for different sparsity levels $n=(2,4,6,8)$; a gradual increase of maximum deviation $\max_{i,t}(|\Delta m_i^z(t)|)$ from 0.03 to 0.37 is observed as sparsity is increased. 
For comparison, we also consider the scenario where instead of locally increasing the sparsity we increase the global Trotter step size $\delta t$ for a uniform Trotterization. 
The maximum magnetization deviation for uniform Trotterization upon varying $\delta t$ is found to be significantly larger, ranging from 0.15 for $\delta t=0.2$ to 1.59 for $\delta t=0.8$, see Figs.~\ref{fig:xy_domainwall_mag_corr_deviation}e--\ref{fig:xy_domainwall_mag_corr_deviation}h.
A similar trend is also observed for the maximum deviation of the magnetic correlations, $\max_{i,t}(|\Delta\chi_{0,i}^{zz}(t)|)$, which ranges from 0.03 to 0.36 for sparse Trotterization with $n$ between 2 and 8 (Figs.~\ref{fig:xy_domainwall_mag_corr_deviation}i--\ref{fig:xy_domainwall_mag_corr_deviation}l) and from 0.15 to 1.57 for uniform Trotterization with $\delta t$ between 0.2 and 0.8 (Figs.~\ref{fig:xy_domainwall_mag_corr_deviation}m--\ref{fig:xy_domainwall_mag_corr_deviation}p).

\begin{figure}
	\includegraphics[width=\linewidth]{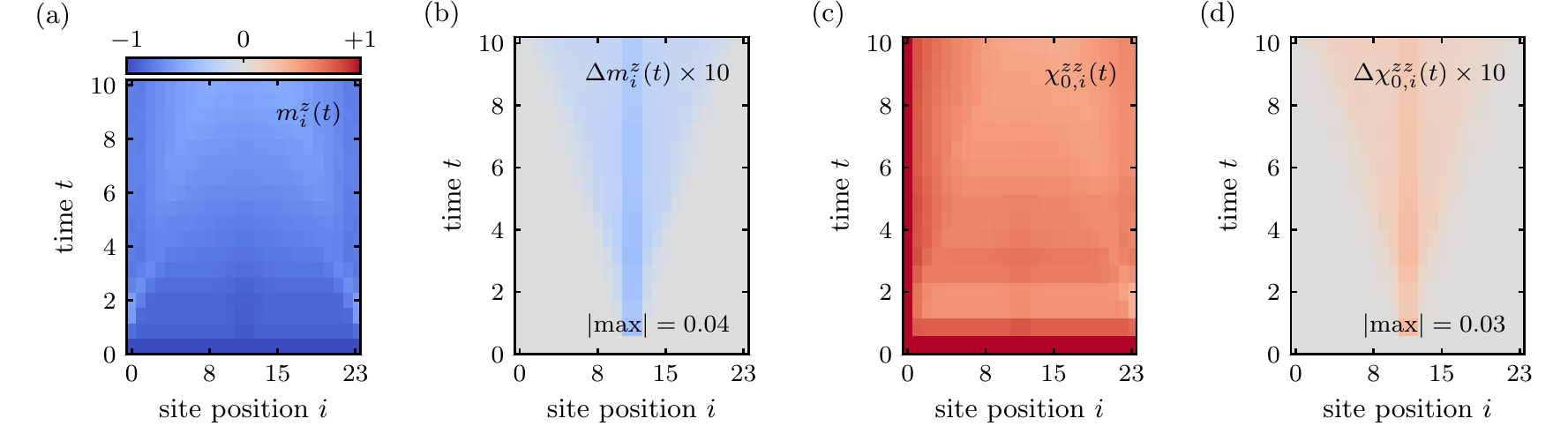}
	\caption{Magnetization and magnetic correlations obtained from sparse Trotterization at $n=6$ for the slow TFI quench model at $h=0.5$. Data is obtained on $k=2$ compute nodes. (a)~Time-dependent local magnetization $m_i^z(t)$ at spin $i$. (b)~Magnetization difference $\Delta m_i^z(t)$ between the state obtained from sparse Trotterization and the reference state $|\psi_\mathrm{ref}(t)\rangle$. The maximum deviation $|\mathrm{max}| = \max_{i,t}(|\Delta m_i^z(t)|)$ is indicated in the panel. (c)~Magnetic correlations $\chi_{0,i}^{zz}(t)$ between the leftmost spin and all remaining spins $i$. (d)~Correlation difference $\Delta \chi_{0,i}^{zz}(t)$ between the state obtained from sparse Trotterization and the reference state. The maximum deviation $|\mathrm{max}| = \max_{i,t}(|\Delta \chi_{0,i}^{zz}(t)|)$ is indicated in the panel. The color bar displayed in panel~(a) applies to all panels. Note that the data in panels (b) and (d) has been rescaled by a factor of 10. Data is obtained for $L=24$ spins.}
	\label{fig:tfi_quench_05_mag_corr}
\end{figure}
The magnetization for the slow TFI quench model, obtained for $k=2$ compute nodes, is shown in Fig.~\ref{fig:tfi_quench_05_mag_corr}a. 
The deviation with respect to the reference state is small; at $n=6$, the maximum deviation is only 0.04. 
For this model, which is initialized in a uniformly magnetized state at the initial time $t=0$, it is particularly visible how the deviations emanate from the center of the spin chain, which is where the sparse bond is located (Fig.~\ref{fig:tfi_quench_05_mag_corr}b). 
A similarly small deviation of 0.03 is observed in the magnetic correlations, see Figs.~\ref{fig:tfi_quench_05_mag_corr}c and \ref{fig:tfi_quench_05_mag_corr}d. 
The maximum deviation that is obtained from uniform Trotterization with increased step size $\delta t$ is much larger. 
For $\delta t=0.6$ the maximum deviation of the magnetization is 0.31 and the maximum deviation of the susceptibility is 0.27. 

\begin{figure}
	\includegraphics[width=\linewidth]{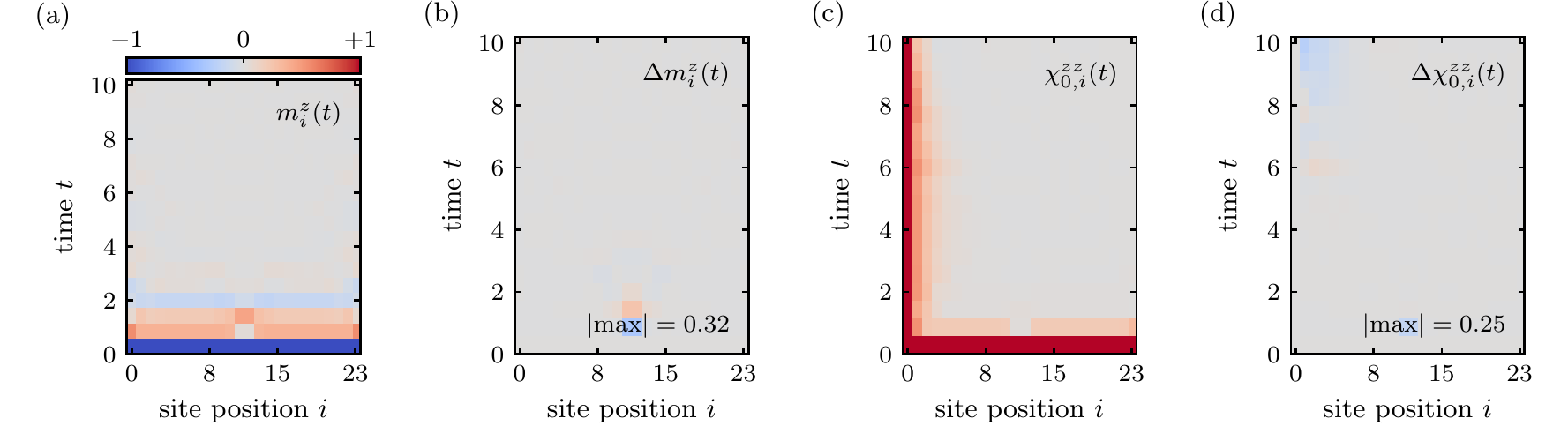}
	\caption{Magnetization and magnetic correlations obtained from sparse Trotterization at $n=6$ for the fast TFI quench model at $h=2.0$. Data is obtained on $k=2$ compute nodes. (a)~Time-dependent local magnetization $m_i^z(t)$ at spin $i$. (b)~Magnetization difference $\Delta m_i^z(t)$ between the state obtained from sparse Trotterization and the reference state $|\psi_\mathrm{ref}(t)\rangle$. The maximum deviation $|\mathrm{max}| = \max_{i,t}(|\Delta m_i^z(t)|)$ is indicated in the panel. (c)~Magnetic correlations $\chi_{0,i}^{zz}(t)$ between the leftmost spin and all remaining spins $i$. (d)~Correlation difference $\Delta \chi_{0,i}^{zz}(t)$ between the state obtained from sparse Trotterization and the reference state. The maximum deviation $|\mathrm{max}| = \max_{i,t}(|\Delta \chi_{0,i}^{zz}(t)|)$ is indicated in the panel. The color bar displayed in panel~(a) applies to all panels. Data is obtained for $L=24$ spins.}
	\label{fig:tfi_quench_20_mag_corr}
\end{figure}
Analogous data for the fast TFI model is displayed in Fig.~\ref{fig:tfi_quench_20_mag_corr}. 
For data obtained from sparse Trotterization at $n=6$, the maximum deviations of magnetization and magnetic correlations are 0.32 and 0.25, respectively. 
For data obtained from uniform Trotterization with $\delta t = 0.6$, the maximum deviations of magnetization and magnetic correlations are 0.67 and 0.79, respectively.

\end{document}